\documentclass{PoS}
\usepackage{graphics}

\title{Kaon and $D$ meson masses with $N_f = 2+1+1$ twisted mass lattice QCD}

\ShortTitle{Kaon and $D$ meson masses with $N_f = 2+1+1$ twisted mass lattice QCD}

\author{ETM~Collaboration: R\'emi~Baron$^a$, Philippe~Boucaud$^{b}$, Jaume~Carbonell$^{c}$, Vincent~Drach$^{c}$, Federico~Farchioni$^{d}$, Gregorio~Herdoiza$^e$, Karl~Jansen$^e$, Chris~Michael$^{f}$, Istv\'an~Montvay$^{g}$, Elisabetta~Pallante$^{h}$, Siebren~Reker$^{h}$, Carsten~Urbach$^{i}$, \speaker{Marc~Wagner}$^{j}$, Urs~Wenger$^{k}$}

\abstract{
We discuss the computation of the kaon and $D$ meson masses in the $N_f = 2+1+1$ twisted mass lattice QCD setup, where explicit heavy flavor and parity breaking occurs at finite lattice spacing. We present three methods suitable in this context and verify their consistency.
}

\FullConference{The XXVIII International Symposium on Lattice Field Theory, Lattice2010\\
		June 14-19, 2010\\
		Villasimius, Italy}

\newcommand{\ltapprox}{\raisebox{-0.5ex}{$\,\stackrel{<}{\scriptstyle\sim}\,$}}

\begin{document}


\section{Introduction}

The European Twisted Mass Collaboration (ETMC) is currently performing large scale simulations with $N_f = 2+1+1$ flavors of dynamical quarks using Wilson twisted mass lattice QCD \cite{Baron:2008xa,Baron:2009zq,Baron:2010bv}.

Particular problems are caused by the non-degenerate strange/charm quark doublet, since besides the usual twisted mass parity breaking the strange and charm quark numbers are not conserved. The latter amounts to contamination of correlation functions by intermediate states with wrong flavor quantum numbers. In particular, correlation functions of charmed mesons and baryons obtain small contributions from similar strange systems, which are significantly lighter and, therefore, inevitably dominate at large temporal separations. In this paper we discuss these problems in the context of the kaon and $D$ meson masses, which are important quantities already at the stage of generating gauge field configurations, when tuning the strange and charm quark masses to their physical values. We present three methods to overcome these problems and demonstrate their consistency.
This work is a summary of a more detailed recent paper \cite{Baron:2010th}.


\section{Simulation setup}

The gauge action is the Iwasaki action \cite{Iwasaki:1985we}. The fermion action is Wilson twisted mass,
\begin{eqnarray}
 & & \hspace{-0.7cm} S_{\mathrm{F},\textrm{\scriptsize light}}[\chi^{(l)},\bar{\chi}^{(l)},U] \ \ = \ \ a^4 \sum_x \bar{\chi}^{(l)}(x) \Big(D_\mathrm{W}(m_0) + i \mu \gamma_5 \tau_3\Big) \chi^{(l)}(x) \\
 & & \hspace{-0.7cm} S_{\mathrm{F},\textrm{\scriptsize heavy}}[\chi^{(h)},\bar{\chi}^{(h)},U] \ \ = \ \ a^4 \sum_x \bar{\chi}^{(h)}(x) \Big(D_\mathrm{W}(m_0) + i \mu_\sigma \gamma_5 \tau_1 + \tau_3 \mu_\delta\Big) \chi^{(h)}(x)
\end{eqnarray}
for the light degenerate up/down doublet \cite{Frezzotti:2000nk} and the heavy non-degenerate strange/charm doublet \cite{Frezzotti:2003xj} respectively, where $D_\mathrm{W}$ denotes the standard Wilson Dirac operator. $\kappa = 1 / (2 m_0 + 8)$ is tuned to maximal twist by requiring $m_{\chi^{(l)}}^\mathrm{PCAC} = 0$, which guarantees automatic $\mathcal{O}(a)$ improvement for physical quantities, e.g.\ the here considered kaon and $D$ meson masses. We also refer to \cite{Chiarappa:2006ae}, where this $N_f = 2+1+1$ twisted mass setup has been pioneered.

All results presented in the following correspond to computations on $1042$ gauge field configurations from ensemble B35.32 \cite{Baron:2010bv} characterized by gauge coupling $\beta = 1.95$, lattice extension $L^3 \times T = 32^3 \times 64$ and bare untwisted and twisted quark masses $\kappa = 0.161240$, $\mu = 0.0035$, $\mu_\sigma = 0.135$ and $\mu_\delta = 0.170$. The corresponding lattice spacing is $a \approx 0.078 \, \textrm{fm}$, the pion mass $m_\mathrm{PS} \approx 318 \, \textrm{MeV}$.


\section{Quantum numbers, physical and twisted basis meson creation operators}

In Wilson twisted mass lattice QCD parity is not a symmetry and the heavy flavors cannot be diagonalized -- both symmetries are broken at $\mathcal{O}(a)$. Consequently, instead of the four QCD heavy-light meson sectors labeled by heavy flavor and parity, $(s,-)$, $(s,+)$, $(c,-)$ and $(c,+)$, there is only a single combined heavy-light meson sector $(s/c,-/+)$ in twisted mass lattice QCD. In contrast to QCD, where the $D$ meson is the lightest state in the $(c,-)$ sector, it is a highly excited state in the combined $(s/c,-/+)$ sector of twisted mass lattice QCD. This in turn causes severe problems for the computation of $m_D$, since the determination of excited states is inherently difficult, when using lattice methods.

To create kaons and $D$ mesons a suitable set of operators is
\begin{eqnarray}
\label{EQN001} \mathcal{O}_j \ \ \in \ \ \Big\{ +i \bar{\chi}^{(d)} \gamma_5 \chi^{(s)} \, , \, -i \bar{\chi}^{(d)} \gamma_5 \chi^{(c)} \, , \, +\bar{\chi}^{(d)} \chi^{(s)} \, , \, -\bar{\chi}^{(d)} \chi^{(c)} \Big\} .
\end{eqnarray}
In the continuum the twisted basis quark fields $\chi$ and their physical basis counterparts $\psi$ are related by the twist rotation
\begin{eqnarray}
\left(\begin{array}{c} \psi^{(u)} \\ \psi^{(d)} \end{array}\right) \ \ = \ \ \exp\Big(i \gamma_5 \tau_3 \omega_l / 2\Big) \left(\begin{array}{c} \chi^{(u)} \\ \chi^{(d)} \end{array}\right) \quad , \quad \left(\begin{array}{c} \psi^{(s)} \\ \psi^{(c)} \end{array}\right) \ \ = \ \ \exp\Big(i \gamma_5 \tau_1 \omega_h / 2\Big) \left(\begin{array}{c} \chi^{(s)} \\ \chi^{(c)} \end{array}\right) .
\end{eqnarray}
$\omega_l$ and $\omega_h$ denote the light and heavy twist angles, which are $\pi/2$ at maximal twist. At finite lattice spacing the procedure is more complicated, since renormalization factors have to be included. The heavy-light meson creation operators (\ref{EQN001}) and the corresponding renormalized operators in the physical basis transform into each other via
\begin{eqnarray}
\label{EQN003} \left(\begin{array}{c} +i \bar{\psi}^{(d)} \gamma_5 \psi^{(s)} \\ -i \bar{\psi}^{(d)} \gamma_5 \psi^{(c)} \\ +\bar{\psi}^{(d)} \psi^{(s)} \\ -\bar{\psi}^{(d)} \psi^{(c)} \end{array}\right)^R \ \ = \ \ 
\underbrace{\left(\begin{array}{cccc}
+c_l c_h & -s_l s_h & -s_l c_h & -c_l s_h \\ 
-s_l s_h & +c_l c_h & -c_l s_h & -s_l c_h \\ 
+s_l c_h & +c_l s_h & +c_l c_h & -s_l s_h \\ 
+c_l s_h & +s_l c_h & -s_l s_h & +c_l c_h
\end{array}\right)}_{= \mathcal{M}(\omega_l,\omega_h)}
\left(\begin{array}{c} +i Z_P \bar{\chi}^{(d)} \gamma_5 \chi^{(s)} \\ -i Z_P \bar{\chi}^{(d)} \gamma_5 \chi^{(c)} \\ +Z_S \bar{\chi}^{(d)} \chi^{(s)} \\ -Z_S \bar{\chi}^{(d)} \chi^{(c)} \end{array}\right) ,
\end{eqnarray}
where $c_x = \cos(\omega_x / 2)$, $s_x = \sin(\omega_x / 2)$ and $Z_P$ and $Z_S$ are operator dependent renormalization constants.

The starting point for the three analysis methods presented in the following sections \ref{SEC001} to \ref{SEC002} are the $4 \times 4$ correlation matrices
\begin{eqnarray}
\label{EQN013} C_{j k}(t) \ \ = \ \ \langle \Omega | \mathcal{O}_j(t) \Big(\mathcal{O}_k(0)\Big)^\dagger | \Omega \rangle
\end{eqnarray}
of spatially extended, i.e.\ APE and Gaussian smeared versions of twisted basis heavy-light meson creation operators (\ref{EQN001}). The smearing parameters have been optimized by minimizing effective masses at small temporal separations (cf.\ \cite{Baron:2010th} for details).


\section{\label{SEC001}Method~1: solving a generalized eigenvalue problem}

\begin{figure}[htb]
\begin{center}
\input{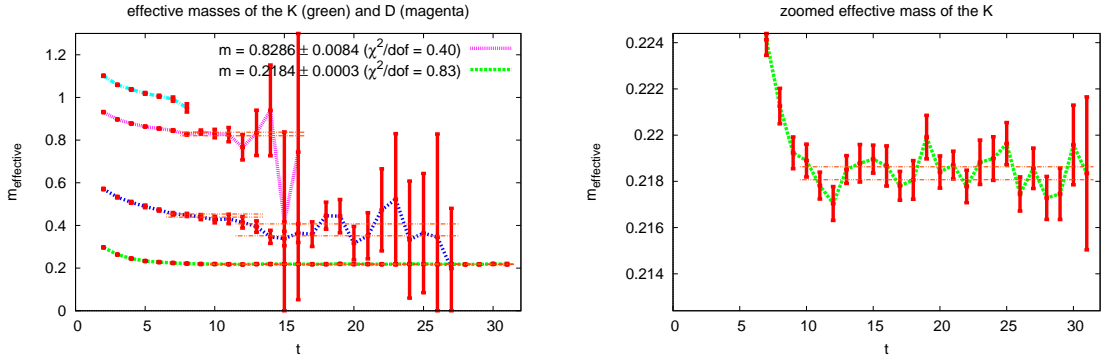}
\caption{\label{FIG001}left: effective masses obtained from the $4 \times 4$ correlation matrix (3.4) by solving the generalized eigenvalue problem (4.1) and (4.2) and corresponding plateau fits; right: zoomed effective mass of the lowest state, which has been identified as the kaon.}
\end{center}
\end{figure}

One possibility to determine $m_K$ and $m_D$ is to solve the generalized eigenvalue problem
\begin{eqnarray}
\label{EQN011} C_{j k}(t) v_j^{(n)}(t,t_0) \ \ = \ \ C_{j k}(t_0)\lambda^{(n)}(t,t_0) v_j^{(n)}(t,t_0)\, ,
\end{eqnarray}
cf.\ e.g.\ \cite{Blossier:2009kd} and references therein. From the eigenvalues $\lambda^{(n)}$ one then computes four effective masses $m_\textrm{\scriptsize effective}^{(n)}$ by solving
\begin{eqnarray}
\label{EQN012} \frac{\lambda^{(n)}(t,t_0)}{\lambda^{(n)}(t+1,t_0)} \ \ = \ \ \frac{\exp(-m_\textrm{\scriptsize effective}^{(n)}(t,t_0) t) + \exp(-m_\textrm{\scriptsize effective}^{(n)}(t,t_0) (T-t))}{\exp(-m_\textrm{\scriptsize effective}^{(n)}(t,t_0) (t+1)) + \exp(-m_\textrm{\scriptsize effective}^{(n)}(t,t_0) (T-(t+1)))}\, ,
\end{eqnarray}
where $T$ is the temporal extension of the lattice and the exponentials $e(-m_\textrm{\scriptsize effective}^{(n)}(t,t_0) (T-t))$ and $e(-m_\textrm{\scriptsize effective}^{(n)}(t,t_0) (T-(t+1)))$ take care of effects due to the temporal periodicity of the lattice. Heavy-light meson masses are finally determined by fitting constants to effective mass plateaus at temporal separations $t \gg 1$. Results are shown in Figure~\ref{FIG001}.

It is well known that for sufficiently large temporal separations $t$ the generalized eigenvalue problem (\ref{EQN011}) and (\ref{EQN012}) yields the lowest states in the sector considered, i.e.\ in our case the four lowest states in the combined $(s/c,-/+)$ sector. The $D$ meson ($m(D) \approx 1868 \, \textrm{MeV}$), however, is not among them. Lighter states include
\begin{itemize}
\item the kaon and its radial excitations, \\ $m(K) \approx 496 \, \textrm{MeV}$, $m(K(1460)) = 1400 \, \textrm{MeV} - 1460 \, \textrm{MeV}$, ...

\item parity partners of the kaon \\ $m(K_0^\ast(800)) = 672(40) \, \textrm{MeV}$, $m(K_0^\ast(1430)) = 1425(50) \, \textrm{MeV}$, ...

\item multi particle states \\ $m(K + \pi)$, $m(K + 2 \times \pi)$, ...
\end{itemize}
At first glance it seems that the $4 \times 4$ correlation matrix (\ref{EQN013}) is not sufficient to determine $m_D$, but that one needs a significantly larger correlation matrix, which is able to resolve all states below the $D$ meson. Note, however, that in the continuum an exact diagonalization of $C_{j k}$ is possible yielding one correlator for each of the four sectors $(s,-)$, $(s,+)$, $(c,-)$, $(c,+)$. Hence the generalized eigenvalue problem would not yield the four lowest masses of the $(s/c,-/+)$ sector, but $m_K$, $m_{(s,+)}$, $m_D$ and $m_{(c,+)}$. At finite lattice spacings such a diagonalization is, of course, only approximately possible. However, discretization artefacts, which are responsible for that, only appear at $\mathcal{O}(a)$ and are thus expected to be small. Therefore, at not too large temporal separations one of the four effective masses should be dominated by the $D$ meson and, consequently, provide an estimate for $m_D$.

\begin{figure}[htb]
\begin{center}
\input{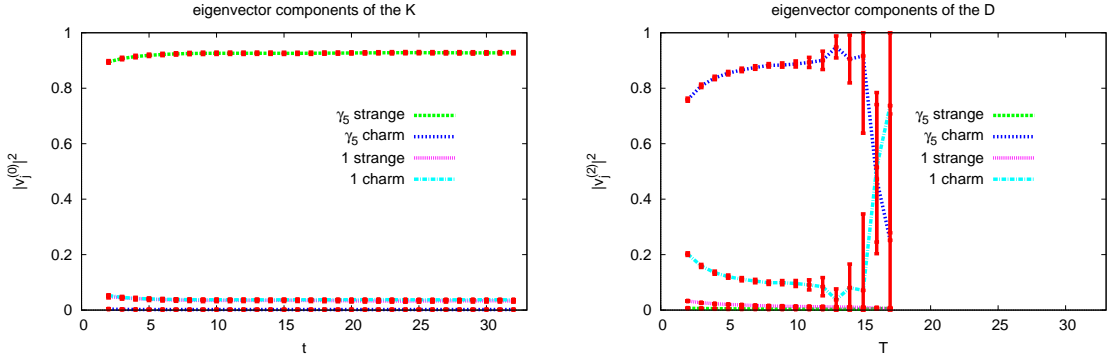}
\caption{\label{FIG002}left: squared pseudo physical basis eigenvector components of the lowest state $|v_j^{(0)}|^2$, which is interpreted as the kaon; right: squared pseudo physical basis eigenvector components of the second excited state $|v_j^{(2)}|^2$, which is interpreted as the $D$ meson.}
\end{center}
\end{figure}

One can check that this is indeed the case by twist rotating the eigenvectors $\mathbf{v}^{(n)}$ to the pseudo physical basis, which is defined by (\ref{EQN003}) with $Z_P = Z_S = 1$. Then one can read off the approximate flavor and parity content of each of the four states corresponding to the four effective mass plateaus. As shown in Figure~\ref{FIG002}, the lowest state is dominated by the physical basis operator $\bar{\psi}^{(d)} \gamma_5 \psi^{(s)}$ and, therefore, interpreted as the kaon, while the second excited state is dominated by $\bar{\psi}^{(d)} \gamma_5 \psi^{(c)}$, i.e.\ corresponds to the $D$ meson.

\section{Method 2: fitting exponentials}

An alternative approach to determine $m_K$ and $m_D$ is to perform a $\chi^2$ minimizing fit of
\begin{eqnarray}
\sum_{n=1}^{N} (a_j^{(n)})^\dagger a_k^{(n)} \Big(\exp\Big(-m_n t\Big) + \exp\Big(-m_n (T-t)\Big)\Big) ,
\end{eqnarray}
i.e.\ of $N$ exponentials to the computed correlation matrix $C_{j k}$ in a suitably chosen window of temporal separations $t$. Notice that $T$ is the temporal extension of the lattice and the second exponential $\exp(-m_n (T-t))$ takes care of effects due to the lattice temporal periodicity. The masses $m_n$ can be interpreted by analyzing the prefactors $a_j^{(n)}$, which is very similar to what has been explained in more detail in the previous section for the eigenvector components provided by the generalized eigenvalue problem.

As before we find that the lowest state is a kaon, while the second excited state is dominated by the physical basis operator $\bar{\psi}^{(d)} \gamma_5 \psi^{(c)}$ and, hence, should correspond to the $D$ meson.


\section{\label{SEC002}Method~3: heavy flavor/parity restoration}

Our third approach is based on the twist rotation of heavy-light meson creation operators (\ref{EQN003}).
In a first step we express the correlation matrix $C_{j k}$ in the physical basis in terms of the twist angles $\omega_l$ and $\omega_h$ and the ratio of renormalization factors $Z_P$ and $Z_S$:
\begin{eqnarray}
\nonumber & & \hspace{-0.7cm} C^{\textrm{\scriptsize physical},R}(t;\omega_l,\omega_h,Z_P / Z_S) \ \ = \\
 & & = \ \ \mathcal{M}(\omega_l,\omega_h)  \textrm{diag}(Z_P,Z_P,Z_S,Z_S) C(t) \textrm{diag}(Z_P,Z_P,Z_S,Z_S) \mathcal{M}^\dagger(\omega_l,\omega_h)\, ,
\end{eqnarray}
where the matrix $\mathcal{M}$ has been defined in (\ref{EQN003}).
We then determine $\omega_l$, $\omega_h$ and $Z_P / Z_S$ by requiring
\begin{eqnarray}
\label{EQN004} C_{j k}^{\textrm{\scriptsize physical},R}(t;\omega_l,\omega_h,Z_P / Z_S)\Big|_{j \neq k} \ \ = \ \ 0 .
\end{eqnarray}
At finite lattice spacing and small temporal separations $t$ this cannot be achieved exactly, because of $\mathcal{O}(a)$ heavy flavor and parity breaking effects and the presence of excited states. However, at sufficiently large $t$, where only the kaon survives, the condition (\ref{EQN004}) can be realized. It amounts to removing any kaon contribution from the diagonal correlators $C_{j j}^{\textrm{\scriptsize physical},R}$, $j \neq (s,-)$.

Finally we analyze the diagonal correlators $C_{j j}^{\textrm{\scriptsize physical},R}$ individually. There is one correlator for each of the four sectors $(s,-)$, $(s,+)$, $(c,-)$ and $(c,+)$. The effective mass plateaus corresponding to the $(s,-)$ and the $(c,-)$ diagonal correlator yield the heavy-light meson masses $m_K$ and $m_D$, respectively.


\section{Conclusions and outlook}

\begin{figure}[htb]
\begin{center}
\input{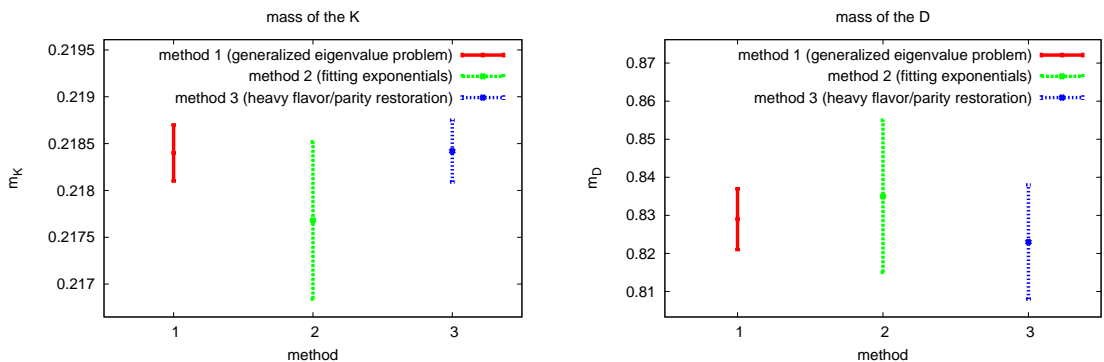}
\caption{\label{FIG003}comparison of kaon and $D$ meson masses determined with the three methods presented.}
\end{center}
\end{figure}

Results for $m_K$ and $m_D$ obtained with our three methods agree within statistical and systematic errors, see Table~\ref{TAB001} and Figure~\ref{FIG003}. For a detailed discussion, of how statistical and systematic errors have been determined, we refer to \cite{Baron:2010th}. We are able to determine $m_K$ in a rigorous way with rather high statistical precision (statistical error $\ltapprox 0.4 \%$). On the other hand all three methods require assumptions, when computing $m_D$, which amount to a systematical error being involved. Nevertheless, the combined statistical and systematical error for $m_D$ is $\ltapprox 2.5 \%$.


\begin{table}[htb]
\begin{center}

\begin{tabular}{|c|c|c|c|}
\hline
 & & & \vspace{-0.40cm} \\
 & method 1 & method 2 & method 3\vspace{-0.40cm} \\
 & & & \\
\hline
 & & & \vspace{-0.40cm} \\
$m_K $ & $0.2184(3)$ & $0.2177(8)$ & $0.2184(3)$ \\
$m_D$ & $0.829(8){\phantom 0}$  & $0.835(20)$   & $0.823(15)$\vspace{-0.40cm} \\
 & & & \\
\hline
\end{tabular}

\caption{\label{TAB001}comparison of kaon and $D$ meson masses determined with the three methods presented.}

\end{center}
\end{table}


Being able to determine $m_K$ and $m_D$ is very important for tuning the strange and charm quark masses to their physical values. For precision charm physics, however, we intend to use a mixed action Osterwalder-Seiler setup \cite{Frezzotti:2004wz}. First steps in this direction are currently under way and have been reported during this conference \cite{CU2010}.


\begin{acknowledgments}

The computer time for this project was made available to us by the John von Neumann-Institute for Computing (NIC) on the JUMP, Juropa and Jugene systems in J\"ulich and apeNEXT system in Zeuthen, BG/P and BG/L in Groningen, by BSC on Mare-Nostrum in Barcelona (\texttt{www.bsc.es}) and by the computer resources made available by CNRS on the BlueGene system at GENCI-IDRIS Grant 2009-052271 and CCIN2P3 in Lyon. We thank these computer centers and their staff for all technical advice and help.

This work has been supported in part by the DFG Sonderforschungsbereich TR9 Computergest\"utzte The\-o\-re\-tische Teilchenphysik and the EU Integrated Infrastructure Initiative Hadron Physics (I3HP) under contract RII3-CT-2004-506078. We also thank the DEISA Consortium (co-funded by the EU, FP6 project 508830) for support within the DEISA Extreme Computing Initiative (\texttt{www.deisa.org}).

\end{acknowledgments}



\end{document}